\documentclass[onecolumn]{emulateapj}

\usepackage{graphicx}
\usepackage{rotating}

\slugcomment{}

\shorttitle{}
\shortauthors{Lacy et al.}

\begin{document}


\title{The {\em Spitzer} mid-infrared AGN survey.  II-the demographics and 
cosmic evolution of the AGN population.}


\author{
M.\ Lacy \altaffilmark{1},
S.E.\ Ridgway\altaffilmark{2}, 
A.\ Sajina\altaffilmark{3},  
A.O.\ Petric\altaffilmark{4},
E.L. Gates \altaffilmark{5},
T. Urrutia\altaffilmark{6},
L.J.\ Storrie-Lombardi\altaffilmark{7}}
\altaffiltext{1}{National Radio Astronomy Observatory, 520 Edgemont Road, Charlottesville, Virginia 22903}
\altaffiltext{2}{National Optical Astronomy Observatory, 950 North Cherry Avenue, Tucson, AZ 85719}
\altaffiltext{3}{Department of Physics and Astronomy, Tufts University, 212 College Avenue, Medford, MA 02155}
\altaffiltext{4}{Gemini Observatory, Northern Operations Center, 670 N. A'ohoku Place, Hilo, HI 96720}
\altaffiltext{5}{UCO/Lick Observatory, P.O.\ Box 85, Mount Hamilton, CA 95140}
\altaffiltext{6}{Leibniz-Institut f\"{u}r Astrophysik Potsdam, An der Sternwarte 16, 14482, Potsdam, Germany}
\altaffiltext{7}{Spitzer Science Center, Caltech, Mail Code 220-6,
Pasadena, CA 91125; mlacy@ipac.caltech.edu,sajina@ipac.caltech.edu, 
lisa@ipac.caltech.edu}

\begin{abstract}


We present luminosity functions derived from a spectroscopic survey of AGN selected 
from {\em Spitzer Space Telescope} imaging surveys. Selection in the
mid-infrared is significantly less affected by dust obscuration. 
We can thus compare the luminosity functions of the obscured
and unobscured AGN in a more reliable fashion than by using optical or X-ray 
data alone. We find that the AGN luminosity function 
can be well described by a broken power-law model in which 
the break luminosity decreases with redshift.
At high redshifts ($z>1.6$), 
we find significantly more AGN at a given 
bolometric luminosity than found by either optical quasar surveys
or hard X-ray surveys. The 
fraction of obscured AGN decreases rapidly with increasing AGN 
luminosity, but, at least at high redshifts, appears to remain at 
$\approx 50$\% even at bolometric luminosities $\sim 10^{14}L_{\odot}$. 
The data support a picture in which the obscured and unobscured 
populations evolve differently, with some evidence that
high luminosity obscured quasars peak in space density at 
a higher redshift than their unobscured counterparts. The amount
of accretion energy in the Universe estimated from this work suggests that
AGN contribute about 12\% to the total radiation intensity of the Universe, 
and a high radiative accretion efficiency $\approx 0.18^{+0.12}_{-0.07}$
is required to match current estimates of the local 
mass density in black holes.

\end{abstract}



\section{Introduction}

The existence of a large population of obscured AGN has been recognized
for as long as AGN have been known. The vast majority of well-studied
obscured objects, however, are local, of relatively 
low bolometric luminosity. The fraction of obscured
AGN at the high (quasar) end of the AGN luminosity function, where the 
population is at cosmological distances has been the subject of debate.
Reliable statistics exist only for the $\approx 10$\% of radio-loud AGN,
where luminous 
radio galaxies are seen to exist in similar numbers to radio-loud
quasars when selected on an isotropic property such as low frequency radio 
emission (e.g.\ Willott et al.\ 2000). The situation for radio-quiet
objects is less certain, and it is only in the last decade that progress
has been made in quantifying the population of dust obscured radio-quiet 
quasars through advances in X-ray, optical and mid-infrared selection 
techniques.

Selection of AGN in the mid-infrared allows the discovery of powerful AGN
and quasars whose optical and soft X-ray emission is hidden by dust (e.g.\
Lacy et al.\ 2004, 2007; Stern et al.\ 2005; 2012; Mart\'{i}nez-Sansigre et al.\
2005; Donley et al.\ 2012; Eisenhardt et al.\ 2012). 
Mid-infrared selection has the powerful characteristic 
that it permits the selection
of samples of AGN 
containing both moderately obscured and unobscured objects of 
similar bolometric luminosities, allowing an estimate to be made of the 
importance of the obscured AGN population to the AGN population as a whole.
Mid-infrared
selection is also a useful
complement to other techniques for finding obscured quasars such as 
hard X-rays (Norman et al.\ 2002; Brusa et al.\ 2010), which tend to be
restricted to smaller fields and thus lower-luminosity objects, and optical
techniques based on narrow-line emission (Zakamska et al.\ 2003; 
Alexandroff et al.\ 2013), which are
restricted to redshift ranges in which strong emission lines fall.

Mid-infared selection has its limitations, for example,
Juneau et al.\ (2013) show that, at $z=0.3-1$ and low luminosities 
($L_{\rm bol}\sim 10^{10}L_{\odot}$),  
both X-ray and mid-infrared selection 
miss a large fraction of AGN that are recovered by the Mass-Excitation 
technique, an optical diagnostic which compares the [O{\sc iii}]/H$\beta$
ratio to the stellar mass (Juneau et al.\ 2011). These
objects are probably of such low luminosity that 
hot dust emission does not stand out against stellar light in  
mid-infrared color-color plots. Similar
incompletenesses to lower luminosity AGN have been noted
by Hickox et al.\ (2009), Park et al. (2010), Mendez et al. (2013) and
Messias et al.\ (2014). In particular, 
Messias et al.\ show that the completeness of the AGN 
selection technique of Lacy et al.\ (2007)
rises from $\approx 50$\% compared to X-ray
selection at X-ray luminosities $L_X=10^{43} {\rm ergs^{-1}}$
to 100\% complete at $L_X=10^{44} {\rm ergs^{-1}}$ (i.e.\ 5$\mu$m 
luminosities $\nu L_{\rm 5mu}\sim 10^{43.6}$ to $10^{44.6} {\rm ergs^{-1}}$).
Nevertheless, mid-infrared selection finds many AGN that are missing from
X-ray surveys even at low luminosities (e.g.\ Park et al.\ 2008), so 
mid-infrared remains a powerful technique for finding AGN.

Quantifying the abundance of the obscured population is important for two 
reasons. First, it affects the argument of Soltan (1982), where the 
total amount of matter accreting onto black holes is compared to the
remnant black hole mass density today. The comparison allows us to estimate
the mean value of the radiative efficiency of accretion, $\eta$, which 
varies from $\approx 0.06$ for a non-rotating (Schwartzshild) black
hole to $\approx 0.3$ for a maximally-rotating Kerr black hole. Previous
estimates suggest $\eta \approx 0.1$ (e.g.\ Hopkins, Richards \& Hernquist 
2007, hereafter H07; Mart\'{i}nez-Sansigre \& Taylor 2009; 
Delvecchio et al.\ 2014), though an analysis of the 
X-ray background (Elvis, Risaliti \& Zamorani 2002) 
suggests a higher bound, $\eta > 0.15$, consistent with 
most black holes spinning and with current AGN censuses being significantly 
incomplete. The second reason is to aid in our understanding of 
unified schemes of AGN. In the standard orientation-based unified scheme (e.g.
Antonucci 1993), the ratio of obscured
to unobscured quasars defines the opening angle of the obscuring torus.
(In a refinement of this model, Elitzur (2012) suggests that the 
clumpy nature of the torus means that on an object-by-object basis the
angle to the line of sight is only a statistical predictor of obscuration,
nevertheless the obscured to unobscured ratio is still related to the torus 
opening angle in a statistical sense.) The
luminosity dependence of the obscured to unobscured
ratio (Lawrence 1991; Simpson 2005; 
Gilli, Comastri \& Hasinger 2007; Lusso et al.\ 2013) 
has been put forward as evidence of the ``receding torus'' model
in which more luminous objects are able to move the inner radius of the
obscuring torus out, reducing the covering factor of the dust. Competing
with this theory, in the case of high luminosity and (especially) high 
redshift quasars is the evolutionary theory, whereby quasars begin life highly
obscured as a result of mergers of dusty galaxies
and eventually break out from their cocoons of dust and gas due to outflows
(Sanders et al.\ 1988; Hopkins et al.\ 2008). In
this case the ratio of obscured to unobscured objects can be used to 
estimate the fraction of its lifetime that 
a quasar exists in the obscured state. 

To date, estimates of the fraction of dust-obscured quasars 
have been affected by either small sample
size (Mart\'{i}nez-Sansigre et al.\ 2005; Lacy et al.\ 2007), a restricted
redshift range based on a requirement that [O{\sc iii}] be detected in an 
optical spectrum (e.g. Reyes et al.\ 2008), or a very specific infrared
color selection (Assef et al.\ 2014). 
All these estimates suggest that
the obscured population of quasars equals or exceeds the unobscured
population in space density, even at high luminosities. To increase the 
sample size available to us, we have undertaken a larger survey (detailed
in Lacy et al.\ 2013; hereafter Paper I). This survey is drawn 
from the 54 deg$^2$ covered by the {\em Spitzer} Wide-area Infrared 
Extragalactic Survey (SWIRE; Lonsdale
et al.\ 2003) and the {\em Spitzer} Extragalatic First Look Survey (XFLS; Lacy et al.\ 2005; Fadda et al.\ 2006). AGN and 
quasars were selected on the basis of their mid-infrared properties 
from several samples flux-limited at 24$\mu$m, with  
flux density limits ranging from $S_{24}=$0.61-9.6mJy.  Further selection based on mid-infrared colors at 3.6-8.0$\mu$m was then used to remove most non-AGN from the survey and 
provide a candidate list of 786 objects for which optical and/or near-infrared
spectroscopy was obtained. Of these, 672 were confirmed spectroscopically 
as AGN or quasars. The remainder either showed no firm evidence for an AGN in 
their spectra, had featureless spectra, or had spectra with only a single
weak emission line. In this, the second paper
in the series, we compare the fractions of obscured and unobscured
quasars in this survey, derive the 
luminosity functions of the different AGN types, and 
compare to AGN surveys in other wavebands. We assume a cosmology with 
$H_0=70 {\rm kms^{-1} Mpc^{-1}}$, $\Omega_M=0.3$ and $\Omega_{\Lambda}=0.7$.

\section{The evolution of the AGN population as seen in the mid-infrared}

\subsection{Photometric data}

We used the optical through near-infrared photometric data tabulated 
in Table 2 of Paper I, combined with 12$\mu$m flux densities from the 
{\em Wide field Infrared Survey Explorer (WISE)} (Wright et al.\ 2010).
The 12$\mu$m flux densities were particularly useful for constraining the 
mid-infrared SEDs of the AGN between the {\em Spitzer} 8$\mu$m and 
24$\mu$m bands. 
We matched the objects in Paper I to the {\em WISE} all-sky survey using a
$3^{''}$ match radius, recovering matches to 896 out of 963 (93\%) of objects.
(Some of those not matched had faint detections in WISE 12$\mu$m from which 
we could estimate flux densities.) These data were used as the basis
of the spectral energy distribution (SED)  fits described in the Appendix.

\subsection{The luminosity-redshift plane}

Figure 1 shows the redshift-luminosity plane for our survey. The 
rest-frame 5$\mu$m luminosity, $L_{5\mu {\rm m}}$ was calculated 
by fitting the spectral energy distribution of each object with 
either a pure type-1 AGN model, or a combination of an AGN and a 
host stellar population, as detailed in the Appendix. 
The presence of infrared emission from the host galaxy
may lead to some subtle selection effects, which are also discussed further 
in the Appendix, though we do not believe they have a significant
effect on our derived luminosity function, at least at $z>0.4$.

The rest-frame
5$\mu$m luminosity of the AGN (without a stellar host contribution) was
then used to calculate $L_{5\mu {\rm m}}$. The 14 samples
from which the survey was drawn, with their differing 24$\mu$m
flux density limits
and areas, account for the relatively wide luminosity range at a given 
redshift (a factor of $\approx 30$), in all but the highest redshift bins.
For the analysis
presented in this paper, we use the ``statistical sample'' of 
Paper I, which comprises the 662 objects 
in a 90\% spectroscopically-complete subsample of the entire survey
(479 of which are confirmed AGN). The ``statistical sample'' was defined by 
exploiting the close correlation of $S_{24}$ and emission line flux density. 
Objects in each of the 14 samples were sorted in order of
decreasing $S_{24}$. The value of $S_{24}$ at which the survey completeness fell
below 90\% was then noted, and objects with $S_{24}$ brighter than this included
in the ``statistical sample''. This procedure prevents the samples becoming 
too biassed towards objects which are easy to obtain spectra for (e.g.\ 
normal type-1 quasars).
As described in Paper I, we classify our AGN as normal, blue type-1 AGN, 
red type-1 AGN (dust reddened, but still showing broad lines in the 
rest-frame optical, i.e.\ with $A_V\sim 1$ towards the AGN), 
type-2 AGN (showing narrow lines only, even in the 
rest-frame optical, and thus with $A_V \stackrel{>}{_{\sim}}5$ towards the
AGN), non-AGN (selected as AGN candidates but showing 
no signs of AGN activity in their optical spectra), stars, and objects 
with featureless spectra. 
Our statistical sample consists of 122 normal type-1 
AGN, 86 red type-1's and 271 type-2's, 118 non-AGN, 23 stars and 42 with 
featureless spectra.

\begin{figure*}
\plotone{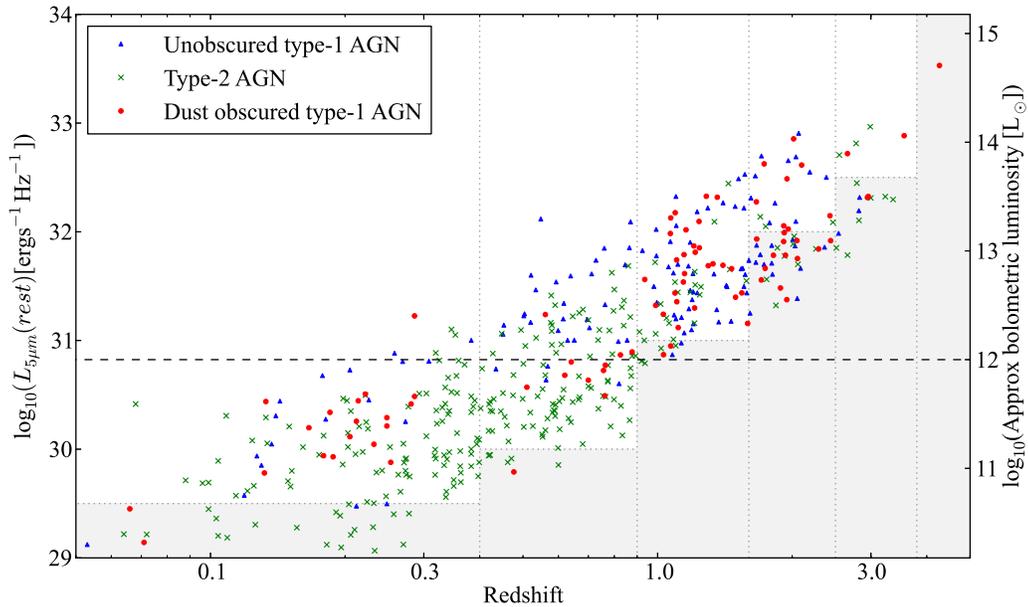}
\caption{Luminosity versus redshift for the survey. The unshaded 
area includes the objects used for the binned analysis of the 
luminosity function, with redshift
bins indicated by dotted lines (all objects were used in the 
maximum-likelihood fit). The dashed horizontal line represents the
approximate division between AGN for Seyfert and quasar luminosities
at a bolometric luminosity of 10$^{12} L_{\odot}$ corresponding to 
${\rm log_{10}}(\nu L_{5 \mu m})\approx 44.6\ {\rm ergs^{-1}}$.}
\end{figure*}

\subsection{The luminosity function}

We performed maximum likelihood fits both to the luminosity functions of the 
AGN population as a whole, and to the individual luminosity functions for the 
normal (blue) type-1s, type-2s and red type-1s. From an initial 
$1/V_a$ estimate of the luminosity function, we noted that, although a 
power-law provided a visually 
reasonable fit to the data, there was a small, but
significant flattening at low luminosities, consistent with the luminosity
function being a broken power-law undergoing luminosity evolution (although
our data do not probe below the break at $z>1$, so the faint-end evolution 
is unconstrained at high redshifts).

We therefore fit a double power-law function of the form:
\begin{equation}
\phi(L,z) = \frac{\phi^{*}}{(L/L^{*})^{\gamma_1^0}+(L/L^{*})^{\gamma_2^0}},
\end{equation}
where $\phi$ is the comoving space density of AGN, $\phi^{*}$ is
the characteristic space density, both in units of comoving Mpc$^{-3}$, 
$L$ is the rest-frame luminosity at 5$\mu$m and $L^{*}$ is the break 
luminosity at 5$\mu$m, 
both in units of ${\rm ergs^{-1}Hz^{-1}}$. The evolution in $L^{*}$ is the usual cubic 
expression (e.g.\ H07):
\begin{equation}
{\rm log_{10}}L^{*}(z)={\rm log_{10}}L^*_0 +  k_1\epsilon + k_2\epsilon^2 + k_3\epsilon^3,
\end{equation}
where
$\epsilon={\rm log}_{10}((1+z)/(1+z_{\rm ref}))$, $L^*_0$ is a free parameter
in the fit, corresponding to the break luminosity at redshift zero, 
$\gamma_1$ is the faint-end slope and  $z_{\rm ref}$ is fixed at $2.5$.
We also fit a model with a fixed faint-end by setting $L^*=L^*_0$ in the
left-hand term in the denominator of Equation 1 and allowing
the bright-end slope, $\gamma_2$ to vary as
\begin{equation}
\gamma_2 = \gamma_2^0+\tau \epsilon .  
\end{equation}

To perform the actual fits, we evaluated the log-likelihood function, 
${\rm ln}\;\mathcal{L}$, in the usual way (Marshall et al.\ 1983), by 
minimizing  $S=-2{\rm ln}\mathcal{L}$ (using the Powell minimization routine
from the scipy.optimize library\footnote{http://scipy.org}):
\begin{equation}
S = -2\sum_{i=1,n_a} {\rm ln}[\phi(L_i,z_i)] + 
2 \sum_{j=1,n_f} \int_{L_{\rm min}}^{L_{\rm max}} \int \phi(L,z) A_j \frac{{\rm d}V}{{\rm d}z}{\rm d}z {\rm d}L 
\end{equation}
where the first term sums over the $n_a$ AGN being considered, and the second 
sums over the $n_f=14$ samples, each of area $A_j$. 
$L_{\rm min}$ is the minimum luminosity detectable at redshift $z$ in a given sample, and
$L_{\rm max}$ the maximum luminosity (for samples where there is an upper flux density limit). 
For the k-corrections in this calculation, mean [5.8]-[8.0] and [8.0]-[24] spectral
indices for each AGN type were measured and applied as appropriate to the AGN type
and the redshift range of the calculation.

The fitting was performed by optimizing the parameters $\gamma_1$, $\gamma_2$, $\tau$, $\phi^*_0$, $k_1,k_2$, $k_3$ 
and ${\rm log_{10}}(L_{0}^{*})$ for AGN with $z>0.05$ and $L>10^{29.5}{\rm ergs^{-1}Hz^{-1}}$. 
We fit all three AGN types separately (normal type-1s, type-2s and red
type-1s). We also fit 
the type-2s and red type-1s together to give 
a luminosity function for the total obscured population. 
In order to assess the possible effects of 
incompleteness, we fit to a set including the objects classified
optically as non-AGN (some fraction of which are, in fact, 
AGN, see Paper I) and also containing the featureless spectrum objects which 
have no redshifts, 
distributed evenly in comoving volume between $z=0.5-5$, and refer to this
as the ``Maximal'' fit. The best-fit parameters are given in Table 1.

For the purposes of comparison with previous work, we also fit a pure power
law of the form
\begin{equation} 
\phi_{\rm pl} = \phi^{*}(L/L^{*}[z])^{\gamma_2}
\end{equation}
where $\gamma_2$ is allowed to vary according to Equation 3. 

To determine whether the luminosity evolution model 
of Equation 1 was a better fit than the fixed faint-end slope model,
or the pure power-law model
of Equation 5, we examined the likelihood ratio of the fits. 
For two fits giving values of $S_1$ and $S_2$ using models with $\nu_1$
and $\nu_2$ degrees of freedom, respectively, $\Delta S = S_1 - S_2$ is
distributed as a $\chi^2$ distribution with $\nu_1-\nu_2$ degrees of 
freedom (e.g.\ Freeman et al. 1998). On this basis, the fixed faint-end
slope model is slightly, but not significantly worse than the 
luminosity evolution model ($\Delta S=1$, with one degree of
freedom). The smallness of this difference is
perhaps to be expected given the poor constraints
on the faint end and its evolution.
$\Delta S$ for the pure power-law 
versus the broken power law model is 30 (Table 1), and the difference in 
degrees of freedom is only one, 
so the broken power-law is formally a much better
fit. 

To make the binned estimates, we binned the data in 
five redshift bins ($0.05<z<0.4$, $0.4<z<0.9$, $0.9<z<1.6$, 
$1.6<z<2.5$ and
$2.5<z<3.8$) assigning a minimum luminosity to each bin at which the bin 
was approximately complete (unshaded region in Figure 1). The lower redshift 
limit was chosen to avoid $z<0.05$, where the Lacy et al.\ (2004) AGN
selection can be confused by the 6.2$\mu$m Polycyclic Aromatic Hydrocarbon
(PAH) emission. The bin spacing is in intervals of 
${\rm log_{10}}(1+z)\approx 0.13$.

The values of the binned estimates were obtained using the formalism of
Hasinger, Miyaji \& Schmidt (2005), where the fitted luminosity function
of Equation (1) is used to predict the number of sources in each
bin, $n_{\rm pred}$ by integrating the luminosity function over the 
luminosity and redshift ranges for each bin for each individual field
in the survey, properly accounting for the upper and lower flux limits 
in each field. The value of the luminosity function at the bin 
centre, $\psi_{\rm mdl}$, is then corrected by the ratio of the 
number of predicted sources in the bin to the actual number observed, 
$n_{\rm obs}$ - i.e.:
 \[ \phi_{\rm b}^{ij}(L_i,z_j) = \frac{\psi_{\rm mdl} n_{\rm obs}}{n_{\rm pred}}. \]  
Although this has the disadvantage of making the 
binned estimated somewhat model dependent, the advantage of this technique is
that it compensates both for the tendency of objects to be found close
to the bottom of a luminosity bin, and at the higher redshift end of
redshift bins where the luminosity function is increasing 
rapidly with redshift (i.e.\ at $z<2$).

\begin{table}
\caption{Best fitting luminosity function parameters}
{\scriptsize
\begin{tabular}{lcccccccccc}
AGN used & $n_a$ &$S$& ${\rm log_{10}}(\phi^*$ & $\gamma_1^0$ & $\gamma_2^0$ & $k_1$& $k_2$ & $k_3$ & ${\rm log_{10}}(L^*_0$\\
         &         &      & $[{\rm Mpc^{-3}}])$        &           &            &      &       &  & ${\rm [ergs^{-1}Hz^{-1}]})$&$\tau$  \\\hline\hline
All      &       446 &10997       & -4.75$\pm0.02$  & 1.07$\pm 0.06$ & 2.48$\pm 0.05$ & 1.05$\pm 0.03$ & -4.71$\pm 0.13$ & -0.034$\pm 0.19$ &31.92$\pm 0.02$&[0]\\
Normal Type-1 &   119 &3374      & -5.18$\pm 0.05$  & 0.25$\pm 0.18$ & 2.68$\pm 0.10$ & 0.537$\pm 0.08$ &-5.48$\pm 0.23$&0.768$\pm 0.76$ &31.99$\pm 0.08$&[0] \\
Type-2   &        243 &6065      & -5.04$\pm 0.03$  & 1.068$\pm 0.12$ & 2.75$\pm 0.10$ & 1.01$\pm 0.04$ &-4.26$\pm 0.14$  &-0.241$\pm 0.25$ & 31.76$\pm 0.02$ &[0] \\
Red Type-1&       84  &2451      & -5.39$\pm 0.05$  & 0.44$\pm 0.20$ & 2.60$\pm 0.12$ & 1.36$\pm 0.09$ &-4.96$\pm 0.27$ &0.232$\pm 0.56$& 32.03$\pm 0.03$ &[0] \\
Type-2+Red Type-1 & 327& 8159 &-4.98$\pm 0.03$ & 1.09 $\pm 0.06$ & 2.61$\pm 0.08$  & 1.165$\pm 0.04$  & -4.45$\pm 0.14$ & -0.23$\pm 0.20$ & 31.91$\pm 0.01$ &[0] \\
Maximal & 588      &13830& -4.66 $\pm 0.02$ &1.38$\pm 0.04$ & 2.47$\pm 0.07$ &1.24$\pm 0.05$ & -4.49$\pm 0.07$  & 0.336 $\pm 0.22$  &  31.94$\pm 0.01$ & [0]\\\hline
Power-law, All& 446 &11027 &-5.28$\pm 0.02$ & - & 2.16$\pm 0.04$ & 1.30$\pm 0.03$ & -4.10$\pm 0.10$ & -0.475$\pm 0.035$ & [32.0] & 1.5$\pm 0.1$\\
Fixed faint end, All&       446 &10998       & -4.91$\pm0.02$  & 0.57$\pm 0.03$ & 2.72$\pm 0.05$ & 1.26$\pm 0.03$ & -4.90$\pm 0.13$ & 0.539$\pm 0.19$ &32.02$\pm 0.02$&1.5$\pm 0.3$\\\hline
\end{tabular}

\noindent
Notes:- $n_a$ is the number of AGN used in the fit.  $\phi^*, \gamma_1, \gamma_2$ and $L^*_0$ are defined in Equation (1), $k_1,k_2$ and $k_3$ in 
Equation (2). Parameters fixed in the fit are indicated with square brackets. To convert these to approximate bolometric luminosity 
functions add ${\rm log_{10}}(8.0 \nu)=14.68$ to ${\rm log_{10}}(L^*_0)$. 
}
\end{table}

\begin{figure*}
\begin{picture}(500,150)
\put(-30,0){\includegraphics[scale=0.5]{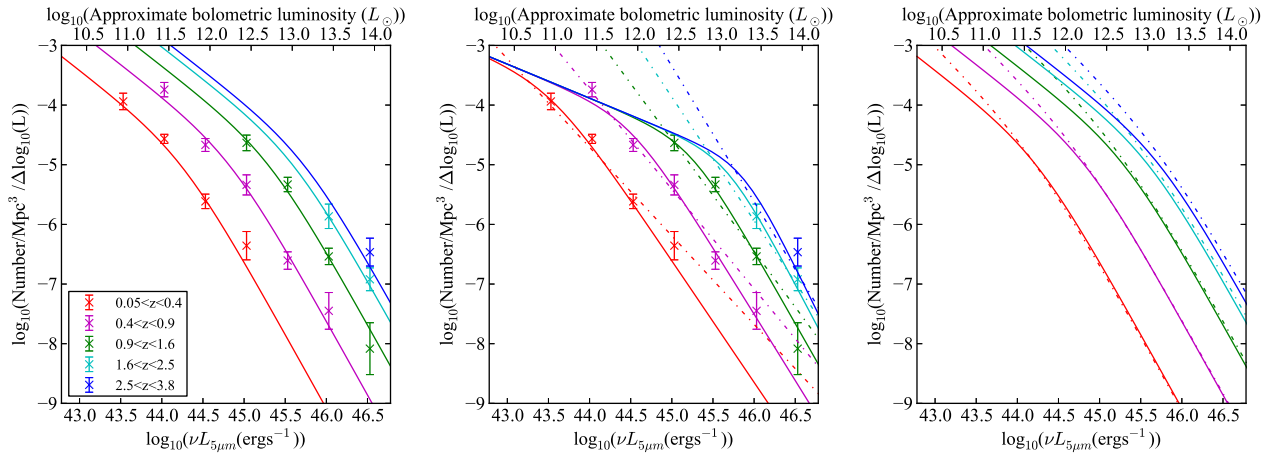}}
\end{picture}
\caption{The evolution of the mid-infrared luminosity function, 
for spectroscopically-confirmed, mid-infrared selected AGN. (a) {\em left}
the best-fit broken power-law model with luminosity evolution
(Equation 1). The points show the binned luminosity function 
and the lines the model. (b) {\em center}
the same, for the fixed faint-end slope model (solid) and 
the  pure power-law model (dot-dashed). (c) {\em right}
the dashed-dot lines show the ``Maximal'' luminosity function, including
objects without AGN spectroscopic confirmation, compared to 
the luminosity evolution model (solid lines).} 
\label{fig:lf_evln}
\end{figure*}

\subsection{Evolution of the AGN number density with redshift}

Figure \ref{fig:lf_evln} 
shows the evolution of the luminosity function of all 
spectroscopically-confirmed {\em Spitzer} 
AGN in our statistical sample with $z>0.05$ and 
$L_{\rm 5 \mu m}>10^{29.5} {\rm ergs^{-1}Hz^{-1}}$. 
The left-hand panel, Figure \ref{fig:lf_evln}(a), shows that the 
luminosity evolution model seems to fit the data well across
most of the range of redshift and luminosities. In the center
panel, (b) we show the alternative fixed
faint-end slope and simple power-law models, both 
of which seem to represent the data reasonably well (though the
single power-law is, statistically, a worse fit than either of the
broken power-law models).
On the right-hand panel (c) we show the ``Maximal'' model
compared to the luminosity evolution one.
Deeper surveys, with AGN identifications 
at $z>1$, are needed to determine the faint end behaviour and
evolution, though Juneau et al.\ (2013) 
show that the faint end of the luminosity function
seems to have evolved little since $z\approx 1$.
The evolution at $z>2$ is relatively poorly
constrained, but, as described in Section 2.6, there is evidence for
a turnover in number densities of high luminosity AGN at $z\approx 2.8$.
AGN at the break luminosity 
contribute most of the luminosity density at a given redshift. 
Our finding that 
the break luminosity is a strong function of redshift, decreasing with 
cosmic time at $z<2.8$ is consistent with the ``downsizing'' scenario
of galaxy evolution (Cowie et al.\ 1996), in the sense that, for a given 
mean Eddington rate, the
characteristic mass of the black holes contributing the bulk of
the AGN luminosity density is decreasing with cosmic time.

When the optically-classified 
``non-AGN'' and blank spectra with randomly-assigned redshifts are added
into the survey to produce the Maximal luminosity function, as shown in 
the right-hand panel of Figure 2 as the dot-dashed line,
we see an increase in the magnitude of the
faint-end slope. At high luminosities, above the break, however, we see 
little more than an increase in the normalization (as expected). 
The larger numbers of low-$z$ AGN is consistent with the population 
of optically-classified
``non-AGN'' being in fact dominated by low luminosity, heavily obscured AGN.
These conclusions must remain speculative, however, 
as it is very difficult to rule out the possibility that the 
some of the ``non-AGN'' are interlopers.

\subsection{Comparison with other AGN luminosity functions}

\begin{figure*}
\plotone{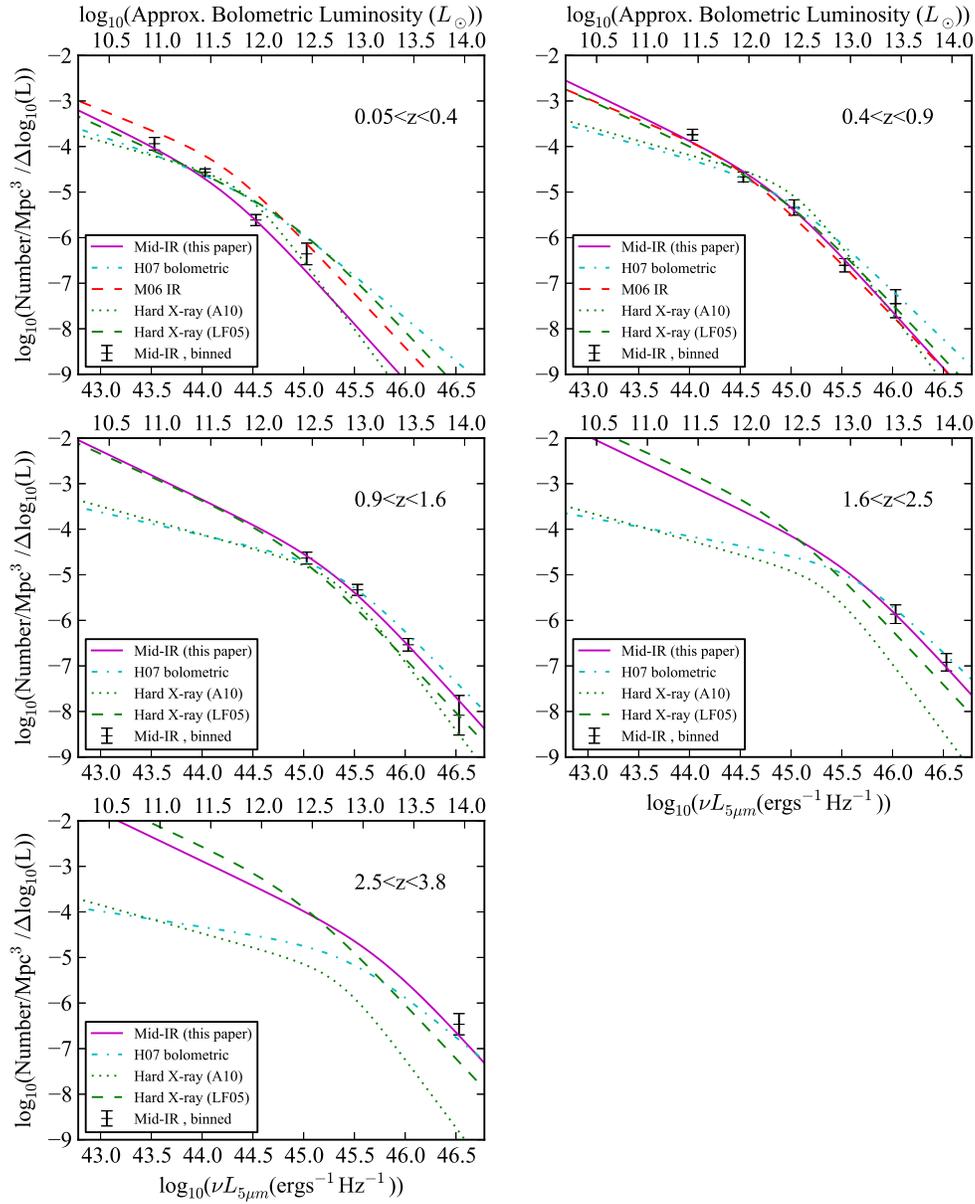}
\caption{Comparison of the mid-infrared and hard X-ray luminosity functions. 
We show both the fit (magenta solid line) and binned (black points with error bars)
versions of our mid-infrared luminosity function, together with the 
Matute et al.\ (2006; M06) mid-infrared luminosity function and the
hard X-ray luminosity functions of Aird et al.\ (2010; A10) and La 
Franca et al.\ (2005; LF05). We also show the  
estimate of the bolometric AGN luminosity function of H07. 
A single, fixed set of bolometric corrections (Table 2) has been 
applied to map all the luminosity functions to 5$\mu$m.}
\label{fig:agn_comp}

\end{figure*}

\begin{figure*}
\plotone{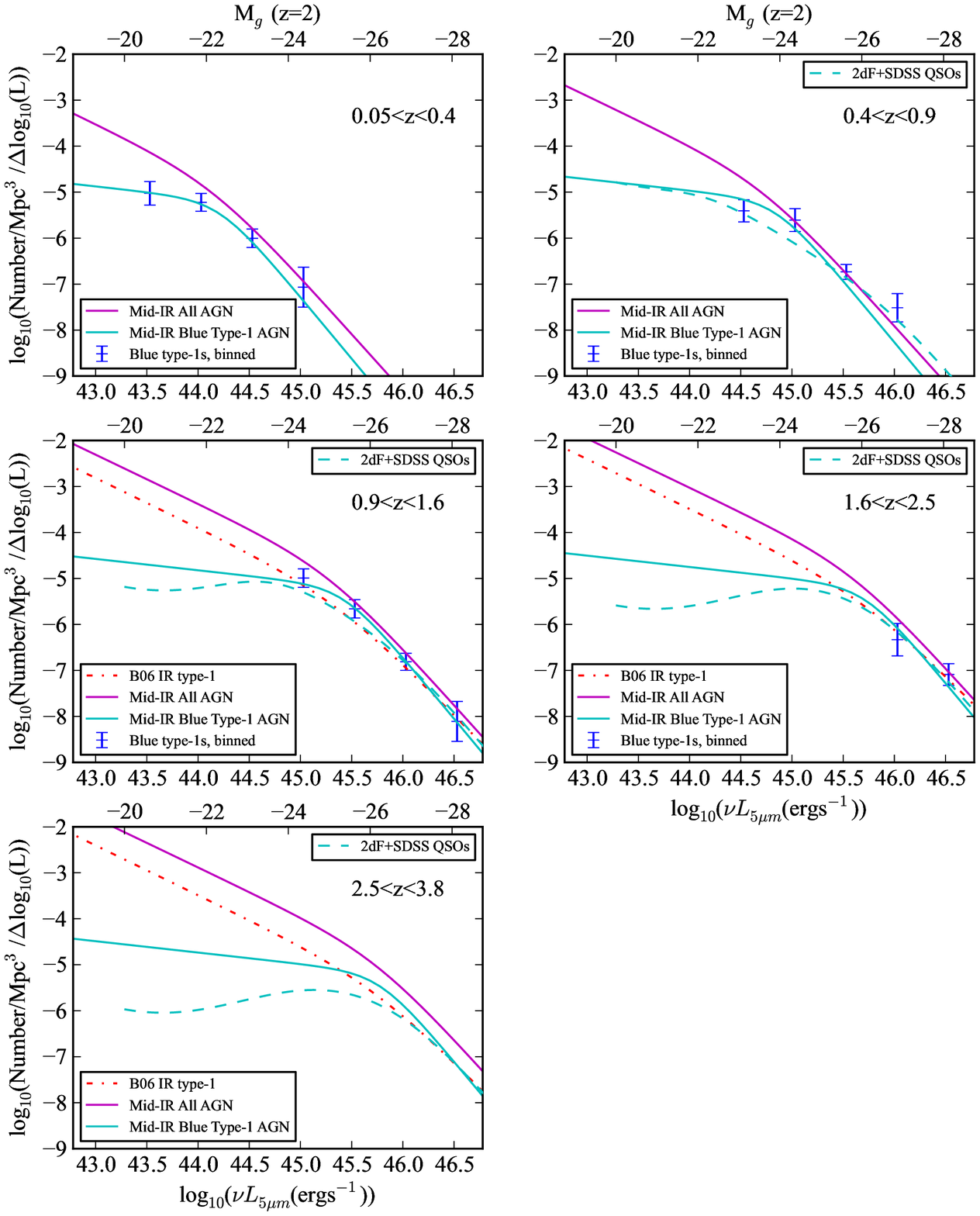}
\caption{Comparison of luminosity functions for normal (blue) type-1
quasars from our survey and the 
2dF and Sloan Digital Sky Survey combined, Croom et al.\ (2009). 
We also plot the luminosity function for mid-infrared selected 
type-1 quasars from Brown et al.\ (2006; B06), which, although 
it contains some reddened quasars was too shallow in the 
optical ($R<21.7$) to find the
bulk of the reddened population.}
\label{fig:type1_comp}

\end{figure*}

In Figure \ref{fig:agn_comp}, we plot a comparison
of our mid-infrared AGN luminosity function with the hard X-ray luminosity 
functions from LaFranca et al.\ (2005) and Aird et al. (2010), 
the mid-infrared luminosity function 
from Matute et al.\ (2006; based on surveys using the 
{\em Infrared Space Observatory}) at $z<0.8$,  
and the estimate of the bolometric 
luminosity function of H07. In Figure \ref{fig:type1_comp}
we plot the normal (blue) type-1 quasar 
luminosity function from our survey, and compare to both the 
combined Sloan Digital Sky Survey (SDSS) and 2-degree Field (2dF) 
quasar surveys luminosity function (Croom et al.\ 2009), and the
mid-infrared luminosity for type-1 AGN from Brown et al.\ (2006) at $z>0.8$.

For both simplicity and ease
of comparison, a single bolometric correction, 
independent of redshift and luminosity, was assumed for each population 
(Table 2). For our mid-infrared AGN, we assume a uniform bolometric correction of 10.
This number was derived as follows. We began with the bolometric correction 
from 5$\mu$m to the total type-1 quasar emission, $C_{\nu}^{'}= 8.0$ 
from Richards et al. (2006).
We then corrected $C_{\nu}^{'}$ for unobscured type-1s to allow for the 
fact that the mid-infrared emission
is likely to be much more isotropic than the optical/UV, and have its origin
in reprocessed UV light originally emitted away from our line of sight. 
We thus derive a ``true'' 
bolometric correction for type-1 objects, $C_{\nu 1} \approx 0.6 C_{\nu}^{'}=5$,
where 0.6 is the typical ratio of the UV/optical to the total quasar emission. 
Considering the type-2s on the other hand, the lower mid-infrared luminosities of 
type-2 objects at a given radio 
luminosity (Paper I, section 6) suggests that type-2 objects have higher 
mid-infrared extinctions than the type-1s. (This is also manifest as a 
steeper 8-24$\mu$m  spectral index for the
type-2s, $\approx 1.4$ compared to $\approx 1.0$ for the type-1s.)
Residual extinction at mid-infrared
wavelengths results in an underestimate of the mid-infrared luminosity for 
the type-2s by a factor $\approx 3$. Thus the effective 
value of the bolometric correction for the type-2 population will be 
$C_{\nu 2}\approx 3 C_{\nu 1}=15$. For simplicity, we average these two
corrections to obtain a value of $C_{\nu}=10$. However, we will return 
to this issue in more detail in future work, where we will undertake a
detailed comparison of type-1 and type-2 AGN SEDs.

At low redshifts ($z<0.3$), we expect to be incomplete due to 
contaminating emission from PAHs of up to $\approx 50$\% (Paper I, section 
2.2), and indeed we find that at $\nu L_{5\mu m}\sim 10^{44-45}$ we 
are below most other AGN luminosity functions in our 
$0.05<z<0.4$ bin, including the mid-infrared 
one of Matute et al.\ (2006). At $0.4<z<0.9$, however,
the luminosity function matches those in other wavebands much better, 
and we see a small excess over the X-ray luminosity function at low
luminosities ($\stackrel{<}{_{\sim}} 10^{44} {\rm ergs^{-1}}$), and 
a very large one over the optical quasar luminosity function.
Considering the blue type-1 objects in our sample 
alone, however, Figure \ref{fig:type1_comp} shows that we agree very well 
with prior estimates of the normal type-1 luminosity function at all redshifts.

As is already well-known (Hasinger 2008; Aird et al.\ 2010), the hard 
X-ray luminosity function peaks at much lower redshift than either the 
mid-infrared or the optical quasar luminosity function. 
This difference between the X-ray and other
luminosity functions has often been 
ascribed to a luminosity-dependent bolometric correction
(e.g.\ Steffen et al.\ 2006; H07, Han et al.\ 2012).
Any redshift-dependent contribution due to absorption effects
has been thought to be low,
though it has long been known that X-ray absorbers are commonly
seen in individual high redshift quasars (Elvis et al.\ 1994b).
The partial correlation analysis of Steffen et al.\ (2006) in particular
seems to indicate a stronger luminosity than redshift dependence on the
ratio between optical and X-ray fluxes, $\alpha_{OX}$ at a 
given optical or X-ray luminosity (though for X-ray luminosities the 
result is only significant at 3$\sigma$), which would translate
into a difference in the bright-end slope of the luminosity function.
However, we find no evidence that a strong luminosity-dependent bolometric 
correction is required based on our comparison of luminosity functions,
and we suspect that the apparently high (13.6$\sigma$) significance
of the anticorrelation $\alpha_{OX}$ on {\em optical} luminosity is a
by-product of luminosity-dependent obscuration effects, i.e.\ 
AGN with lower luminosities tend to have higher gas columns in the
X-ray, increasing the magnitude of $\alpha_{OX}$. In Figure 3, we see that, at $0.4<z<1.6$, 
the mid-infrared, optical quasar, and hard X-ray
luminosity functions are very similar at the high luminosity 
($\nu L_{5\mu m} >10^{45} {\rm ergs^{-1}}$) end of the 
luminosity function. 
Instead of a difference in the slope of the 
AGN luminosity functions between infrared and X-ray, which is what 
we would expect if there was a strong luminosity dependence to the 
bolometric correction, 
what we see is a fall-off in the $z>1.6$ hard
X-ray AGN population at all luminosities (particularly when the 
Aird et al.\ luminosity function is considered, though this luminosity 
function is thought to be somewhat low at $z>3$ (Kalfountzou et al.\ 2014)). 
This suggests that the difference
in the luminosity functions is predominantly a redshift effect.
The lack of a luminosity-dependent bolometric correction in the X-ray  is 
particularly illustrated when the bolometric luminosity function of 
H07 (which assumes a luminosity-dependent bolometric correction)
is plotted in Figure 3, when we see a significant overprediction of the 
space density of $\nu L_{5\mu m} >10^{45.5}{\rm ergs^{-1}}$ AGN
at $z\stackrel{<}{_{\sim}}2$.

\begin{table}
\caption{Assumed bolometric corrections}
{\scriptsize
\begin{tabular}{lcl}
Wavelength             & Correction & Reference\\\hline
Mid-infrared (5$\mu$m) &  10.0              & This paper\\
Hard X-ray (2-10keV)   &  40.0              & Elvis et al.\ (1994a)\\
Optical/UV (1670${\rm \AA}$ rest)& 4.5              & Richards et al.\ (2006)\\\hline
\end{tabular}
}
\medskip
\end{table}

\subsection{Evolution by type of AGN}

\begin{figure}

\plotone{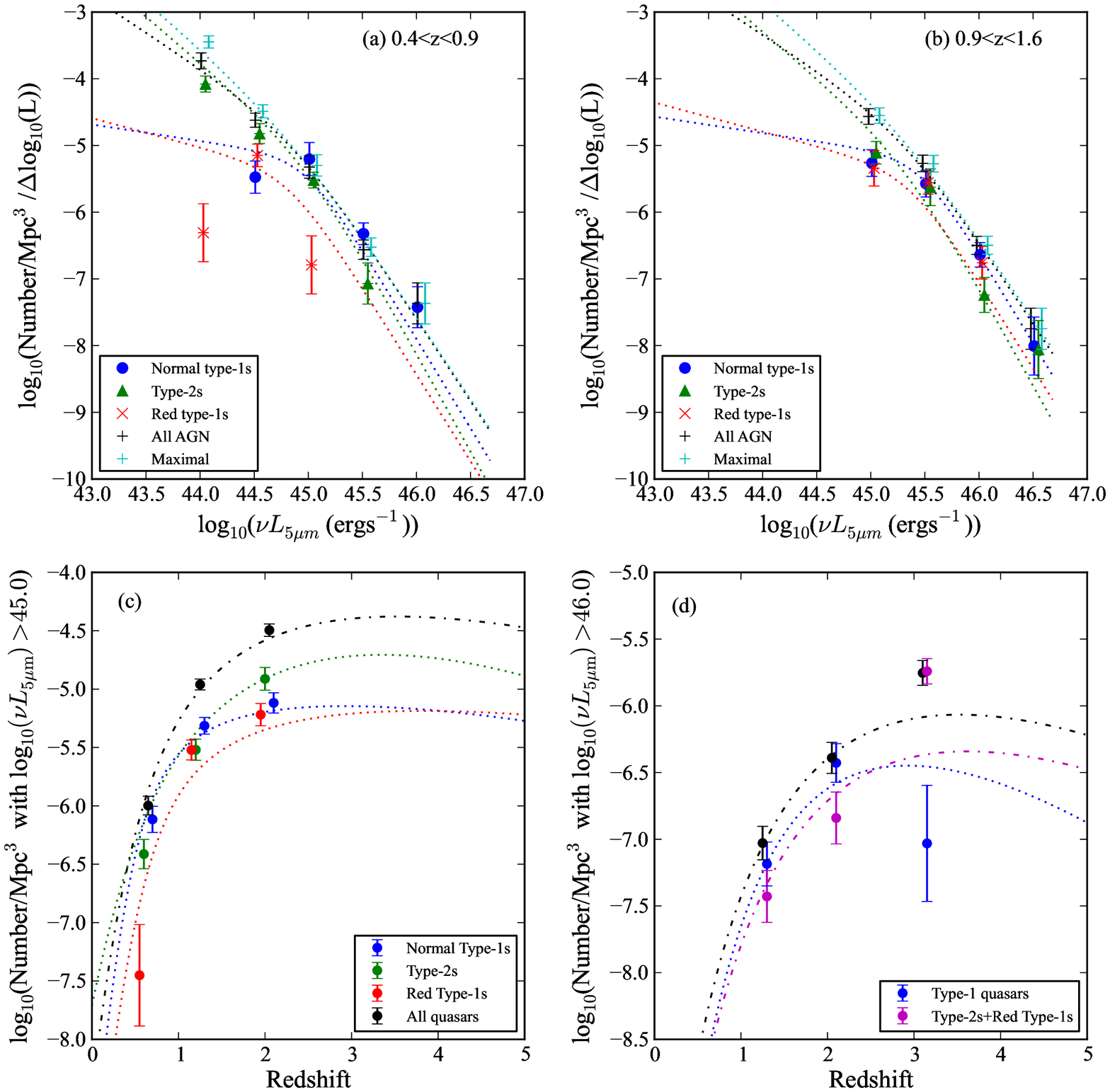}
\caption{Evolution by type of AGN. The top two panels show the differences
in the luminosity functions by type (and also the ``Maximal'' luminosity
function) for two of the best-sampled redshift
bins in the survey. The lower two panels show the number densities of 
highly-luminous quasars ($\nu L_{5\mu m} >10^{45}$ and $10^{46} {\rm erg s^{-1}}$, 
corresponding
to a bolometric luminosity $\approx 10^{12.2}$ and $10^{13.2} L_{\odot}$, respectively) as a function of 
redshift.
The colored dotted lines show the integrated luminosity functions by type, 
and the black dot-dashed lines that for the total AGN population. In the
lower right plot, only the sum of all obscured AGN, the type 1 AGN, and the total of both populations
are shown as magenta, blue and black lines, respectively.}
\label{fig:type_evln}
\end{figure}

Figure \ref{fig:type_evln} 
shows the evolution of the mid-infrared AGN population split up by 
type. The top two panels, (a) and (b), show the luminosity
functions at mean redshifts of 0.65 and 1.25, respectively, for
the different types of AGN, with the plotted points showing the
binned estimates. The black dotted line indicates
the fit for all confirmed AGN, with the cyan line that for the 
Maximal model. Compared to the type-1 luminosity functions (shown in blue
and red, for normal and red type-1s, respectively), the
type-2 luminosity function (green) has a much steeper faint-end slope. 
In the lower two panels, (c) and (d), the number density
of AGN above luminosities of $\nu L_{5 \mu m}=10^{45}$ and 
$10^{46} {\rm ergs^{-1}}$, respectively, are plotted
against redshift. Up to $z\approx 2$ all the populations 
shown (all AGN, blue and red type-1s, type-2s and the combination
of red type-1s and type-2s, representing the obscured population) 
show a similar evolution. The blue type-1 population shows the same evolution as 
optically-selected quasars, with a peak in number density at $z\approx 2.8$, 
but the obscured quasar population (magenta line in panel 
(d)) shows a peak in number density that
is broader and shifted to earlier epochs ($z>3$). 
This is consistent with 
some theoretical predictions (Fanidakis et al.\ 2012), 
results from studies of the radio galaxy population 
(Jarvis et al.\ 2001), and a study of the infrared luminosity 
function of quasars from the 
Sloan Digital Sky Survey (Vardanyan, Weedman \& Sargsyan 2014).
Our models are fairly poorly constrained beyond the peak, however, with only 13 objects 
of any type at $z>2.8$, so the reality of the difference at high redshifts
may be affected by small number statistics. 
Nevertheless, we have performed some tests to investigate
whether the difference in the evolution we see might be statistically significant. We take 
for a null hypothesis that the evolution of obscured AGN (type-2 and red type-1s) and the 
unobscured (type-1) population is the same. We then compare the 
$S$ values for fits which fix the evolution of the obscured population to that of the
unobscured one, and vice-versa. Fixing the values of 
the evolution parameters $k_1,k_2$ and $k_3$
for the obscured population to those fit for the unobscured population, and
allowing the remainder of the parameters to be fit results in a difference
in the $S$ values of 17, which, with the three degrees of freedom from 
fixing the $k_1-k_3$ parameter values allows us to eliminate the null hypothesis 
that the evolution is the same at $>99.9$\% confidence. Forcing the
evolution terms for the unobscured population to be the same as those for
the obscured one results in a less significant 
$\Delta S =9.5$, $\approx 97$\% confidence. We also compared to the 
evolution of all AGN. As expected, fixing the evolution of the 
obscured AGN to match that of all AGN resulted in no significant difference,
but the difference from forcing the type-1 evolution to match that of all AGN 
resulted in ruling out the null hypothesis at the 95\% confidence level.

In summary, we find that the evolution of the obscured and unobscured
AGN populations are significantly different, at $>99.9$\% confidence using
our strongest statistical test. The most obvious difference in the 
luminosity function evolution is at high redshifts, where the 
number density of high luminosity obscured quasars appears to peak at a 
higher redshift than that of the unobscured population. Small number 
statistics, however, mean this interpretation should be treated with caution.

\subsection{The obscured fraction as a function of luminosity and redshift}

\begin{figure}
\plotone{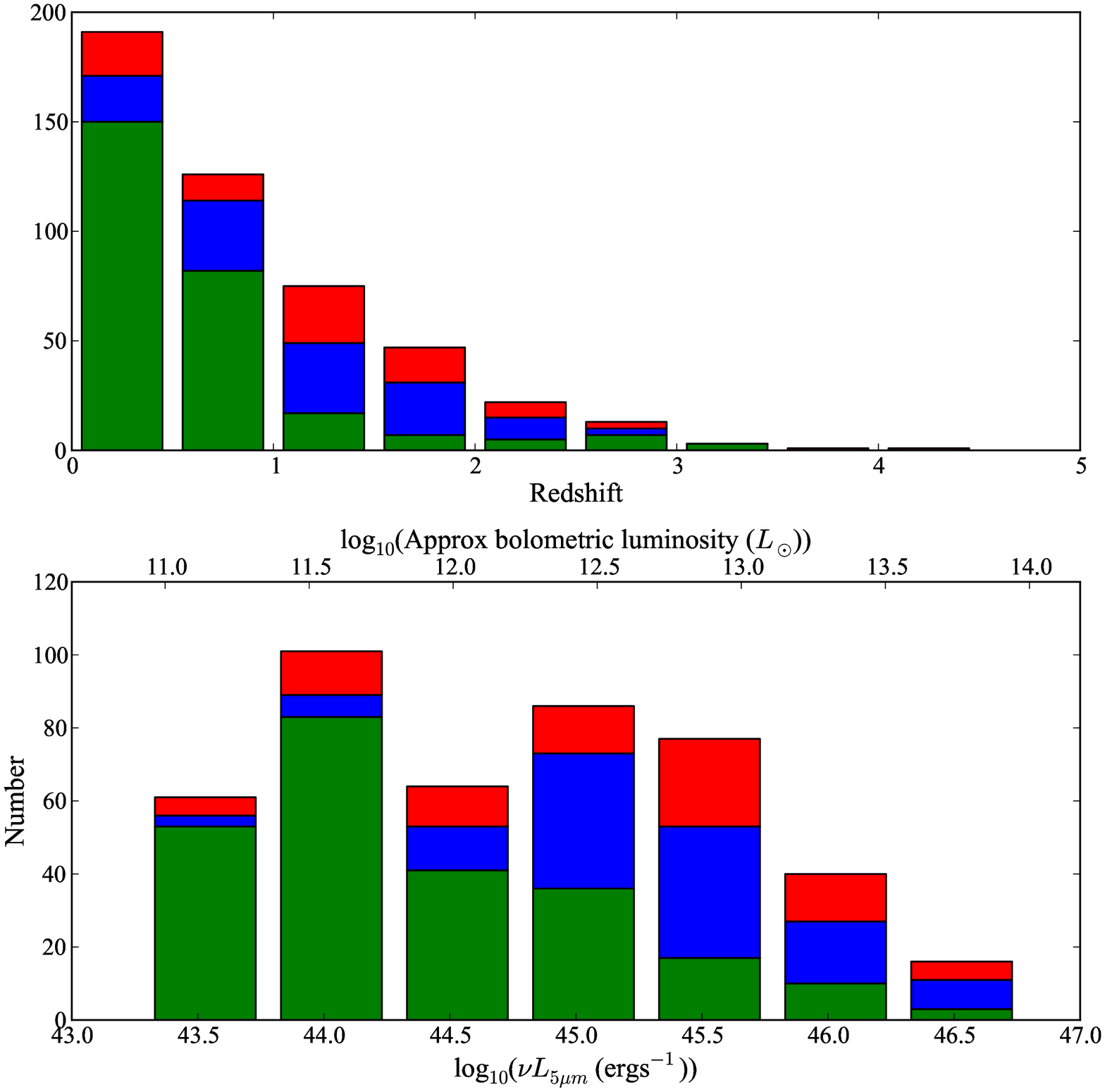}
\caption{The survey objects binned by luminosity. Blue indicates
normal type-1 objects, reddened type-1 objects are shown in red, and 
type-2 objects in green.}
\label{fig:type_hist}
\end{figure}

\begin{figure}
\plotone{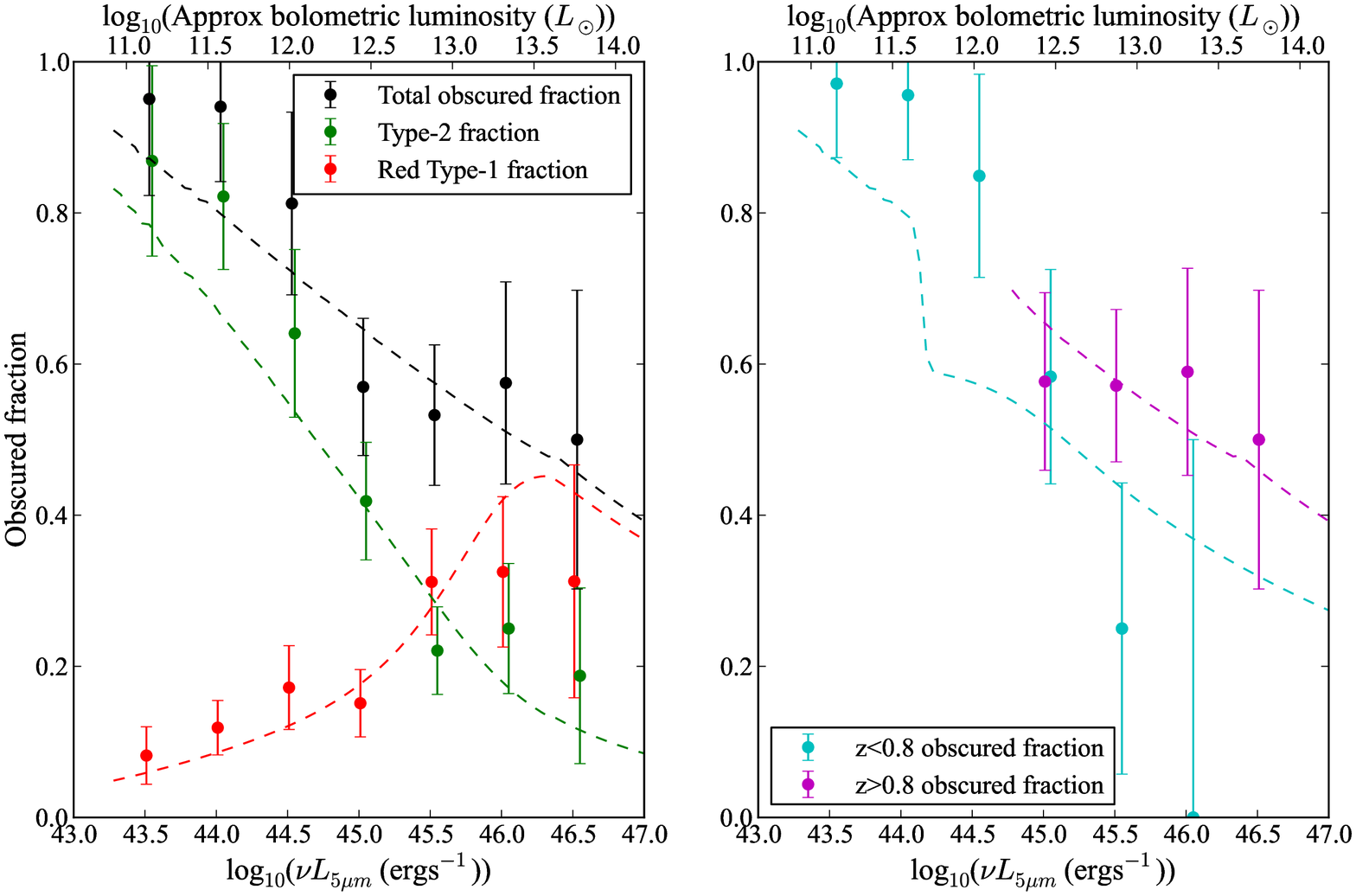}
\caption{The left-hand plot shows the overall obscured 
fraction of AGN (type-2s plus red type-1s) 
declining rapidly as a function of luminosity.
We also show separately the fraction of heavily 
obscured, type-2 objects and lightly obscured, type-1 objects. The
right-hand plot shows the obscured fraction split by redshift into
low ($z<0.8$ and high ($z>0.8$) bins. The measured ratios are
shown as points with errorbars, the ratios of the luminosity 
functions of each type to 
the overall AGN luminosity function (both luminosity functions
integrated over the range in redshift that a given luminosity could
be detected in our survey) are shown as dashed lines.}
\label{fig:obsc_frac}
\end{figure}

Figure \ref{fig:type_hist} shows histograms of number of objects as a function of redshift and luminosity,
split up by type. Figure \ref{fig:obsc_frac} 
plots the fraction of obscured objects as a function of both 
luminosity and redshift with the measured points plotted
with error bars and the luminosity function models plotted as dashed 
lines. These plots show that the fraction 
of obscured objects varies from $\approx 90$\% for 
Seyfert galaxies to $\approx 50$\% at the highest quasar luminosities. 
In addition, there is evidence that the nature of the obscuration in the 
survey seems to change as a function of redshift and/or luminosity. 
At low luminosities and redshifts, most
of the obscured objects are highly obscured, type-2, objects, 
but at high luminosities and redshifts the
obscured population is dominated by less obscured red type-1 
objects. The right-hand panel suggests that, despite the
change in the nature of the obscuration, the overall obscured
fraction is not a strong function of redshift when comparing bins 
at $z<0.8$ and $z>0.8$, and indeed, as discussed in Section 2.6 below, 
it is not until $z>2$ that we see a possible difference.
Of course selection effects may well be playing
a role here as heavily obscured objects become progressively harder to 
identify at high redshifts even in samples selected in the mid-infrared.
As we discuss in Paper I, however, we believe such selection effects are
not having a severe effect on our sample, even at $z>3$.

\subsection{The AGN luminosity density and the black hole mass density}

\begin{figure*}
\begin{picture}(600,180)
\plotone{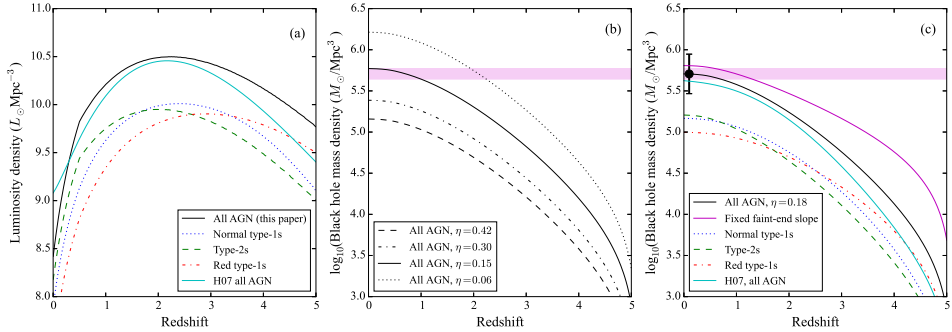}
\end{picture}
\caption{(a) Luminosity density as a function of redshift and AGN type,
based on the luminosity evolution models of Table 1, 
(b) the black hole mass density as a function of the mean 
radiative efficiency of 
the black holes, $\eta$, using the luminosity evolution model for all
AGN, and 
(c) cumulative black hole mass density as a function of redshift  
and AGN type (assuming our best fit value of $\eta=0.18$). In all plots,
the luminosity function has been extrapolated a factor of three below
the survey limit at a given redshift.
The cyan curves in plots (a) and (c) shows the bolometric luminosity 
function of H07, and the magenta curve in (c) shows the fixed-faint end model
from this paper. The magenta stripe in (b) and (c) indicates
the range of the local black hole mass density value estimated by 
Graham \& Driver (2007).}
\label{fig:bh_density}

\end{figure*}

Figure \ref{fig:bh_density} shows the AGN bolometric luminosity density  
and corresponding accreted black hole mass density since $z=5$. Panel (a) 
shows the luminosity density as a function of redshift, with dusty type-1
objects dominating at high redshifts, normal type-1s at $z\sim 2$ and 
type-2 objects at $z<1$. As our best fit luminosity evolution model has a steep faint-end
slope, we need to assume a low luminosity cutoff at high luminosities in order to 
prevent an excessive amount of AGN luminosity density from being estimated. 
We assume, as a fiducial model, that the 
fitted form of the luminosity function we have adopted can be extrapolated a factor
of three below the current limit of the survey. This results in luminosity densities
in line with those we obtain from our alternative 
fixed faint-end slope model. With this assumption, 
we find more AGN than previous surveys, 
and integrating to the present day (e.g.\ Mart\'{i}nez-Sansigre \& Taylor
2009) indicates
a total energy, $U$, emitted by all AGN over the 
history of the Universe of ${\rm log_{10}} (U/{\rm erg\; Mpc^{-3})} = 
37.5 \pm 0.2 $, where the lower bound 
comes from cutting off the faint-end of the luminosity function at 
the luminosity corresponding to the flux limit of the deepest sample in our 
survey (0.61mJy at 24$\mu$m), and the upper bound from extrapolating the 
luminosity function to ten times fainter than this limit.
To obtain the observed energy density today due to AGN, $u$, we need 
to divide the integrand by $(1+z)$ (Soltan 1982),
resulting in a value of $u=2.8\pm 1.2\times 10^{-15} {\rm erg s^{-1} cm^{-3}}$,
or a radiation intensity $b=uc/4\pi=6.7\pm 2.9 {\rm nW sr^{-1}}$,
within the range of $b=3.5-8 {\rm nW sr^{-1}}$ estimated from the 
X-ray background, and suggesting that 
$\approx 12$\% of the total radiation intensity
of the Universe is contributed by AGN (Elvis et al.\ 2002).
We therefore need a radiative efficiency of accretion, 
$\eta$ at the high end of previous estimates to match the 
local mass density in black holes
($\eta \approx 0.15$, Elvis et al.\ (2002); $\eta \approx 0.06$, 
Mart\'{i}nez-Sansigre \& Taylor 2009) (Figure 8 (b), (c)).  
We use the compendium of local black hole mass density estimates 
of Graham \& Driver (2007), with a mean of $10^{5.7}M_{\odot} {\rm Mpc^{-3}}$ 
to estimate a best fit value of $\eta=0.18^{+0.12}_{-0.07}$.

As a function of cosmic time, the evolution of the 
black hole mass density shows a very rapid rise 
with at $z>3$, then
continues to rise, but at a smaller rate. Fifty percent of the current 
mass density has accreted
by $z\approx 1.8$. At $z>3$, we predict that most black hole mass 
accretion occurs in dusty type-1 AGN, this switches to normal 
type-1 quasars at $z\sim 2$, and to type-2 AGN at $z<1$ when powerful 
quasars become very rare.

\section{Discussion}

We present evidence that, when largely free from the effects of 
dust obscuration, the AGN luminosity function for actively-accreting 
black holes seems to be 
well represented by a double power-law with a break that
is a strong function of luminosity, though the form
of the faint end remains unconstrained at $z\stackrel{>}{_{\sim}}1$. 
Downsizing, in the sense that the characteristic AGN
luminosity at which most of the AGN luminosity density is contributed 
is a strongly increasing function of redshift,
is clearly present at $z<2$. 
Luminosity 
functions are often difficult to interpret in terms of physical models; 
however, at low luminosities, we know that most 
local AGN are not triggered 
by major merger events (e.g.\ Kocevski et al.\ 2012). At high
luminosities, especially in the obscured population, major mergers
may indeed play the dominant role (e.g.\ Urrutia et al.\ 2008; 
Treister et al.\ 2012; Hopkins, Kocevski \& Bundy 2013). We therefore
speculate that the increase in the break luminosity with redshift
is driven by the increased frequency of major mergers 
of gas-rich galaxies at high redshifts.

A comparison of the mid-infrared luminosity function with those from 
other wavelengths indicates that we are seeing many more dust-obscured
AGN than the normal type-1 AGN population, and more than the
hard X-ray population, particularly at high
redshifts. We find that the evolution of the obscured and the unobscured
AGN population are significantly different, and 
we see tentative evidence for an earlier peak in the
space density of obscured quasars, consistent with that seen in the 
radio-loud population. 
Our study then is consistent with the evolutionary
explanation for obscured AGN, whereby AGN begin heavily obscured and outflows
clear out material from the line of sight to the AGN. 

Our new luminosity function has allowed us to improve the constraints
on the luminosity density produced by AGN, and on the integral of this
over cosmic time. Our overall AGN
luminosity density agrees well with the range predicted 
from the X-ray background
(Elvis et al.\ 2002), suggesting that we have
indeed accounted for the vast majority of AGN in the Universe, 
and that about 12\% of the luminosity density of the Universe is contributed
by AGN. When compared to the mass density in
compact central objects in galaxies
today, this luminosity density allows an estimate of 
the mean radiative efficiency of accretion onto them. This number 
constitutes one of the best constraints we have on the nature of the
compact central objects in galaxies. Although it is usually assumed
that these are black holes, there are some theoretical alternatives, 
such as gravastars, that can deliver comparable radiative efficiencies
(Harko, Kov\'{a}cs \& Lobo 2009). However, it seems likely that the
radiative efficiencies of such objects are $\eta <0.15$ (Bambi 2012), 
comparable to the current limit from the X-ray background 
(Elvis et al.\ 2002), but below our best estimate of $\eta=0.18$. 
Significant systematic uncertainties remain, however. 
AGN missing from our survey, because they are not 
confirmable in other wavebands (such as the 
``non-AGN'' objects in this paper) or are missing from the mid-infrared
selection criteria (such as low luminosity AGN) will 
only serve to increase our estimate of the radiative efficiency.  
However, there
remain uncertainties on the local black hole mass density, the appropriate
bolometric corrections to apply to both obscured and unobscured objects, 
and the low luminosity end of the luminosity function at high redshifts 
that could result in either an increase or 
a lowering of our estimate for $\eta$. The
bolometric correction uncertainty
can be addressed by using a full multiwavelength dataset, 
including archival Herschel data, to obtain accurate template SEDs for
the obscured AGN population. Improved depth can be 
obtained by using more sophisticated AGN selection techniques
based on the complete infrared SED to exploit the full depth
of SWIRE to select AGN up to four times fainter than the
current limit, and photometric redshifts based
on new deep near-infrared surveys in the SWIRE fields
(SERVS, Mauduit et al.\ 2012; VIDEO, Jarvis et al.\ 2013,
and UKIDSS DXS, Lawrence et al.\ 2007) to obtain redshifts
for these AGN, whose emission lines can be too faint to detect
even using 8m class telescopes.

\acknowledgments

We thank the referee whose comments significantly improved the paper.
The National Radio Astronomy Observatory is a facility of the National 
Science Foundation operated under cooperative agreement by Associated 
Universities, Inc.

\appendix

\section{Optical through mid-infrared SED fitting}

In order to disentangle the contribution of host galaxy light
from our AGN fluxes, we performed SED fitting on the 
photometry described in Section 2.1, using the 
modeling techniques of Sajina et al.\ (2006, 2012, see also Lacy et al.\
2007b and Hiner et al.\ 2011). We fit the SEDs at wavelengths from optical
through 24$\mu$m using the photometry described in Section 2.1. 
For unobscured and red type-1s we fit a pure quasar template from Richards 
et al.\ (2006) (with an SMC reddening law applied in the case of the 
dusty red quasars, and a host galaxy stellar population added in 
a few low luminosity objects). These appear to work fairly well, 
apart from the 1-5$\mu$m region where the AGN SED is known to 
vary as a function of luminosity (e.g.\ Gallagher et al.\ 2007).
For type-2s, we used 
the mid-infrared AGN template of Sajina et al.\ (2012), 
and a host galaxy stellar population, modeled as a single stellar population
from Bruzual \& Charlot (2003) with an age of 0.1,0.64, 1.4 or 5 Gyr picked
to best fit the optical SED. 0.64 Gyr was used as the default in cases
where the optical/near-IR data were too sparse to distinguish between models
(the vast majority of cases), however, 
21 were fit with with a 1.4Gyr population, 23 with a 100Myr population, and
ten with a 5Gyr population. All the stellar population models had similar
behavior in the near-infrared region of the spectrum, however, so the exact
choice of stellar age does not affect the AGN flux density at rest-frame
5$\mu$m by more than a few percent.
Examples of our fits are shown in Figure 9. Full SED fits through the 
far-infrared will be discussed in detail in a future paper.

The SED fitting highlights some ways in which our sample
may be subject to subtle selection effects due to the prescence 
of mid-infrared emission from the host galaxy, particularly
at low luminosities and redshifts. Our use of flux 
limits at 24$\mu$m ensures that photospheric emission from stars
contibutes negligibly ($\stackrel{<}{_{\sim}}1$\% )
to the flux density at the selection frequency. However, there is
the potential for some objects at low redshifts to have their 
24$\mu$m flux densities boosted by emission from warm dust 
due to star formation. Although this will not contribute
significantly to the AGN flux density estimated at 5$\mu$m, it may
lead to objects being included in the sample that would otherwise have
fallen below the flux density limit at 24$\mu$m. However, these same
objects are, to a greater extent, selected against in the survey as their
7.7$\mu$m PAH features will lift them out of the 
AGN selection region at $z<0.4$ (Section 2.5). As the slope of the
warm dust SED is very steep at wavelengths between $\sim 10\mu$m and
24$\mu$m, the contribution of warm dust emission to the 
observed 24$\mu$m flux density at higher redshifts is likely
to be less than a few percent.

\begin{figure*}

\plotone{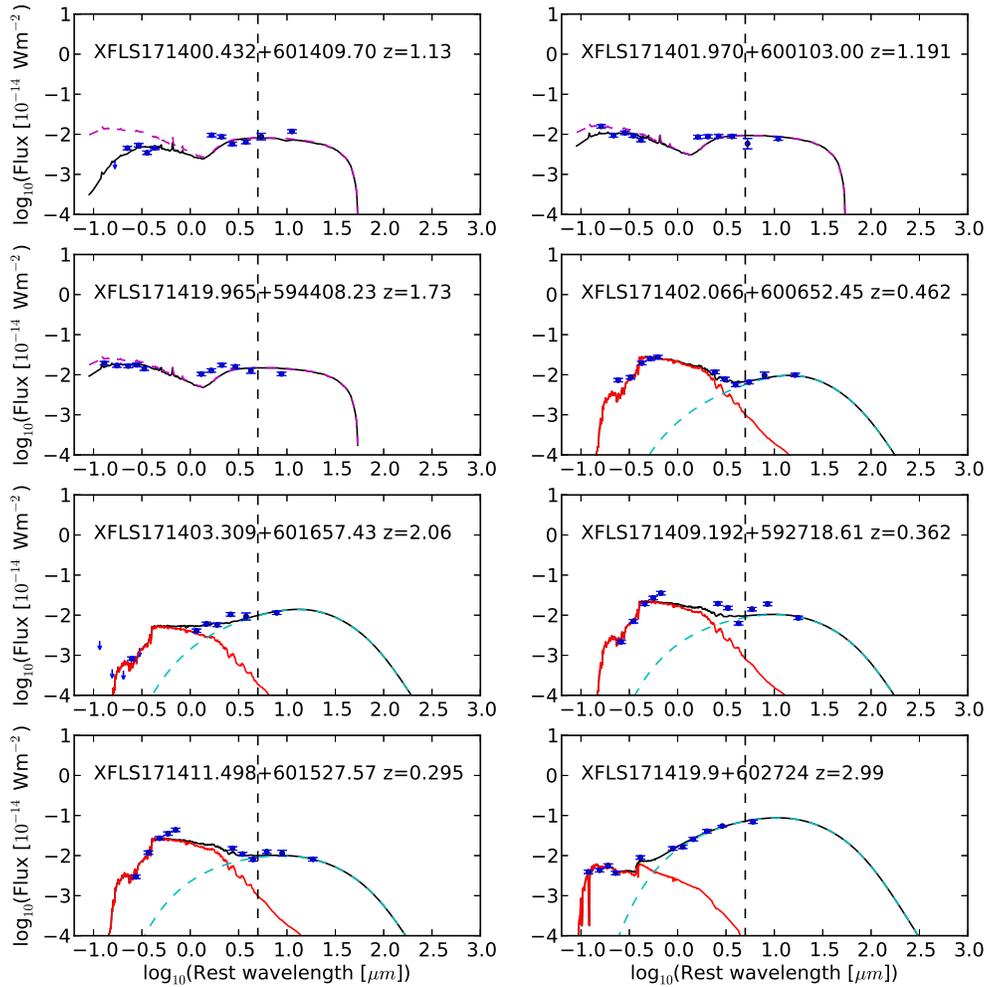}

\caption{Examples of eight SED fits. The overall fit (including reddening) 
is shown as the black line. In the case of type-1 objects, the dashed magenta line
shows the unreddened quasar template. In the case of type-2 objects, or
objects not classified as AGN in their optical spectra, the red line
is the best-fit stellar population, and the dashed cyan line the 
mid-infrared AGN template fit. The dashed vertical line is at a rest
wavelength of 5$\mu$m, indicating the wavelength at which we measure
the AGN flux density.}

\end{figure*}

\end{document}